\documentclass{article}
\usepackage[utf8]{inputenc}
\usepackage{authblk}
\usepackage{natbib}
\usepackage{graphicx}
\usepackage{diagbox}
\usepackage{caption}
\usepackage{subcaption}
\usepackage{float}
\usepackage{color,soul}
\usepackage{url}
\usepackage{todonotes}
\usepackage{lineno}
\providecommand{\keywords}[1]
{
  \small	
  \textbf{\textit{Keywords---}} #1
}

\title{Calibrating the dynamic Huff model for business analysis using location big data \thanks{The final version of this manuscript can be found in Transactions in GIS.}}
\author[1]{Yunlei Liang}
\author[1]{Song Gao \thanks{Corresponding author email: song.gao@wisc.edu}}
\author[1]{Yuxin Cai}
\author[2]{Natasha Zhang Foutz}
\author[3]{Lei Wu}
\affil[1]{Geospatial Data Science Lab, Department of Geography, University of Wisconsin-Madison, Madison, WI, USA}
\affil[2]{School of of Commerce, University of Virginia, VA, USA}
\affil[3]{Data Science, WeWork, Palo Alto, CA, USA}
\date{Jan 30, 2020}

\begin{document}
\maketitle
\begin{abstract}
The Huff model has been widely used in location-based business analysis for delineating a trading area containing potential customers to a store. Calibrating the Huff model and its extensions requires empirical location visit data. Many studies rely on labor-intensive surveys. With the increasing availability of mobile devices, users in location-based platforms share rich multimedia information about their locations in a fine spatiotemporal resolution, which offers opportunities for business intelligence. In this research, we present a time-aware dynamic Huff model (T-Huff) for location-based market share analysis and calibrate this model using large-scale store visit patterns based on mobile phone location data across ten most populated U.S. cities. By comparing the hourly visit patterns of two types of stores, we demonstrate that the calibrated T-Huff model is more accurate than the original Huff model in predicting the market share of different types of business (e.g., supermarkets vs. department stores) over time. We also identify the regional variability where people in large metropolitan areas with a well-developed transit system show less sensitivity to long-distance visits. In addition, several socioeconomic and demographic factors (e.g., median household income) that potentially affect people's visit decisions are examined and summarized.

\end{abstract}\hspace{10pt}
\keywords{business analytics, Huff model, big data, spatial interaction}

\section{Introduction}
“Location, location, and location!” The location information is a key component in business intelligence and implementation of crucial, revenue-generating marketing strategies, such as location-based advertisement and services \citep{negash2008business,fan2015demystifying,gao2017mobilegis,huang2018location}. With the increasingly use of social media, smart devices, and mobile apps, users share rich multimedia information about their locations and associated activities, such as working, shopping, or dining, in a granular spatio-temporal resolution with unprecedented breadth, depth, and scale. Those location-based profiles provide invaluable sources of information for various business analytics and recommendation systems. 

While the original Huff model \citep{huff1964defining} and its subsequent extensions have been used to understand a brand's trade area, they are largely static. The availability of granular spatio-temporal mobility data has permitted the examination of the dynamics of customer mobility patterns at the individual level. For instance, several studies have examined the effects of sampling locations on calibrating the original Huff model to delineate trade areas using mobile phone data \citep{lu2017exploring} and social media data \citep{wang2016evaluating}. In addition, at the aggregated level, stores' trade areas dynamic shift as driven by various potential factors, such as seasonality, marketing strategies, geo-socio-economic changes surrounding the stores, or individuals' dynamic behaviors. Consider, predicting where or which type of location that an individual would visit is also about when the individual is regarding the temporal dynamics of human mobility patterns \citep{ye2011you,yuan2012extracting,gao2015spatio,mckenzie2015poi,yang2016understanding,tu2017coupling}, social relations \citep{shi2015human,xu2017friends}, and semantic configuration and regional variability for temporal signatures of points of interest (POIs) \citep{mckenzie2015regional,liu2020visualizing}. Customers may exhibit different temporal visit preferences to different types of stores, resulting in dynamically shifting trade areas of these stores over different time periods. For example, the grocery stores usually have more daily visits over the weekends than on weekdays. The traditional Huff model can only have one static estimation to each store, which ignores the potential temporal information. However, the temporal dynamics of POIs' visits in cities can be more accurately captured by using large-scale mobile phone location tracking data, facilitating the calibration of a ``dynamic" Huff model to better represent dynamic trade areas at a more granular temporal scale. 

To this end, in this research, we present a time-aware dynamic Huff model (T-Huff) for business location analysis by augmenting the original Huff Model with a dynamic element to capture the time-varying probability of store visitation at the individual customer level. At the aggregated level, the resulting dynamic market share model is calibrated by large-scale store visits based on mobile phone location tracking data. We aim to answer the following two research questions (RQ):
\par RQ 1: How accurate is the dynamic Huff model in predicting the market share of different types of business (e.g., supermarkets vs. department stores) over time? 

\par RQ 2: How do spatial and socioeconomic factors determine the customer choice of particular store visits? Are there any regional variability for store visits in different cities? 

The contribution of this research is threefold: (1) we propose a dynamic Huff model (T-Huff) to estimate hourly store visits from a particular neighborhood over time; (2) by using large-scale individual-level POI visit data across ten most populated U.S. cities, we calibrate the T-Huff model parameters using the technique of particle swarm optimization (PSO) and find that the T-Huff model outperforms the static Huff model when estimating store temporal visits, although regional variability persists; (3) we demonstrate that various factors, such as distance, neighborhood total population, and socioeconomic variables (e.g., median household income, race, and ethnicity diversity), entail distinct influence on the store visits across categories and brands.

The remaining of the manuscript unfolds as follows: we review the relevant literature in section \ref{sec:literature} before introducing the formulations of the original and dynamic Huff models in section \ref{sec:methods}. Then we present the data and study area under analysis in section \ref{sec:data} and report the key empirical findings of the proposed model for three top chain-store brands across ten U.S. cities and discuss the broader implications in section \ref{sec:results}. Finally, we draw conclusions and share our vision for future work in section \ref{sec:conclusion}.

\section{Literature Review} \label{sec:literature}
There is a rich tradition in the marketing literature to study store traffic and its driving factors. For example, \cite{hutchinson1940traffic} used surveys to measure the amount of traffic passing by a Morgantown, WV shoe store and identified 13 factors which could impact sales, including seasonal variations, weather, general business conditions, purchasing power, special location factors, price levels and competition. \cite{bennett1944consumer} studied the out-of-town buying habits in a Maryland town located between D.C. and Baltimore; and found that in many categories purchases were made out-of-town in Baltimore because the survey respondents preferred the proximity of the stores in Baltimore as compared with their town in terms of shopping convenience. To understand how opening a branch store will impact the parent store’s performance, \cite{blankertz1951consumer} conducted a study revealing that branch and parent stores do not attract separate customer groups; rather, both drew trade from substantially the same group; nearby customers in the ``buffer" area between branch and parent traveled most frequently inward to the downtown shopping center despite the greater travel and time involved. 

Then \cite{huff1964defining} defined a trading area as “a geographically delineated region, containing potential customers for whom there exists a probability greater than zero of their purchasing a given class of products or services offered for sale by a particular firm or by a particular agglomeration of firms.” \cite{stanley1976image} further suggested a series of modifications to the Huff model to evaluate the potential of prospective retail store locations. 

This literature further evolved into more sophisticated location analysis, for instance, to advise store site selections. \cite{rosenbloom1976trade} reported on the formation and application of the retail strategy matrix that incorporated three relevant factors: a store's geography, consumer demand, and the area's heterogeneity for identifying and selecting new trade areas for retail stores; and also suggested methods that can be used to adjust the merchandise of existing retail outlets to their trade locations. \cite{ghosh1983formulating} presented a procedure to help retailers formulate a strategic location plan in a dynamic environment, which involved a model for assessing site desirability, a criterion for selecting among alternative sites, and a heuristic to facilitate the computational procedure. More broadly, \cite{grether1983regional} called for more regional-spatial analysis in marketing research. 

The development in this area has also propelled methodological innovation. For example, \cite{fotheringham1988note} proposed a competing destinations model to study hierarchical spatial choices of stores and showed its superior performance as compared to other choice models, such as the nested logit model. \cite{donthu1989note} used kernel density estimation to estimate the spatial distribution of customers in a market and showed how a density-based product positioning methodology may be applied to site selection for a new or re-located store or distribution center. \cite{rust1995capturing} accounted for geographically localized misspecification errors in store choice models with omitted variables that can be correlated with geographic location. They showed that spatial non-stationarity of the model parameters may also be expressed as an instance of omitted variables and therefore be addressed using their method. 

The more recent literature in this domain has focused on location-based competition among stores or chains. A positive association between the number of larger stores and the number and size of smaller stores is reported, implying a mutually beneficial relationship among different types of retailers rather than an overwhelming competitive advantage for larger stores \citep{miller1999effects}. \cite{vitorino2012empirical} used a strategic model of entry to study the store configurations of all U.S. regional shopping centers and to quantify the magnitude of inter-store spillovers. The author showed that consistent with the agglomeration and clustering theories, firms may have incentives to co-locate despite potential business stealing effects; and that the firms' negative and positive strategic effects help predict both how many firms can operate profitably in a given market and the firm-type configurations. In the context of retail outlet locations in the fast food industry, both McDonald's and Burger King were shown better off avoiding close location competition if the market area is large enough; but in small market areas, McDonald's would prefer to be located together with Burger King; in contrast, Burger King's profits always increased with greater differentiation \citep{thomadsen2007product}. Regarding customer's location awareness, \cite{jiang2019solving} calibrated the Huff model with social media data and found that the customers far from the existing retail agglomerations may be more sensitive to the distance. 

Furthermore, studying price competition among (gasoline) retailers conditional on geographic locations, \cite{chan2007econometric} found that consumers were willing to travel up to a mile for a savings of \$.03 per liter. \cite{talukdar2008cost} found the price differentials between wealthy and poor neighborhoods to be 10\%-15\% for everyday items. Even after controlling for the store size and competition, prices were found to be 2\%--5\% higher in poor areas, which was explained by access to cars that acted as a key determinant of consumer’ price search patterns.

In sum, the original Huff model and its subsequent extensions have been widely used to model a brand or a store 's trade area and to predict customer visit probability, but they are largely static. Recent research by \cite{mckenzie_et_al_cosit2017} demonstrated that thematic regions can be represented dynamically using place-type specific temporal patterns. Customers have different temporal visit preferences to different types of stores. It requires a dynamic model to better capture the spatiotemporal characteristics of customers' store visit behaviors.

\section{Methods} \label{sec:methods}
\subsection{The original Huff model}
The Huff model was introduced in order to provide a probabilistic analysis of shopping center trade area, which is a region containing potential customers for a store \citep{huff1963probabilistic,huff1964defining}. The identification of a trade area for a store is crucial as the business owner can estimate how many potential customers will visit this store within this region, and therefore, being able to predict the market sales of this store among competing businesses.

The Huff model proposes that there are two major factors affecting the number of potential customers of a store, which is essentially a gravity-based spatial interaction model. The first one is the merchandise offerings, namely, the ability of the store to fulfill the customers' needs \citep{huff1963probabilistic}. This is also called the attractiveness of a store. If a store has a great number of items, it is able to attract more customers even from distant regions. The other factor is the travel time or travel distance to visit a store. As the expense of traveling to that store increases, the willingness of visiting that store could be significantly reduced \citep{huff1963probabilistic}. 

Based on those two factors, the probability of one customer traveling to a given store can be denoted as follows:
\begin{equation}
P_{ij} = \frac{\frac{S_{j}^\alpha}{D_{ij}^\beta}}{\sum_{j=1}^{n}\frac{S_{j}^\alpha}{D_{ij}^\beta}}
\end{equation}
where $P_{ij}$ is the probability of a customer $i$ visiting a store $j$; $S_j$ is the attractiveness of the store $j$; $D_{ij}$ is the physical distance between the customer $i$ and the store $j$. $n$ indicates there are $n$ stores that a customer $i$ can visit. The parameters $\alpha$ and $\beta$ are used to reflect the effects of attractiveness and the distance on the model.

\subsection{A time-aware dynamic Huff model (T-Huff)}
\begin{figure}[H] 
	\centering
	\includegraphics[width=0.98\textwidth]{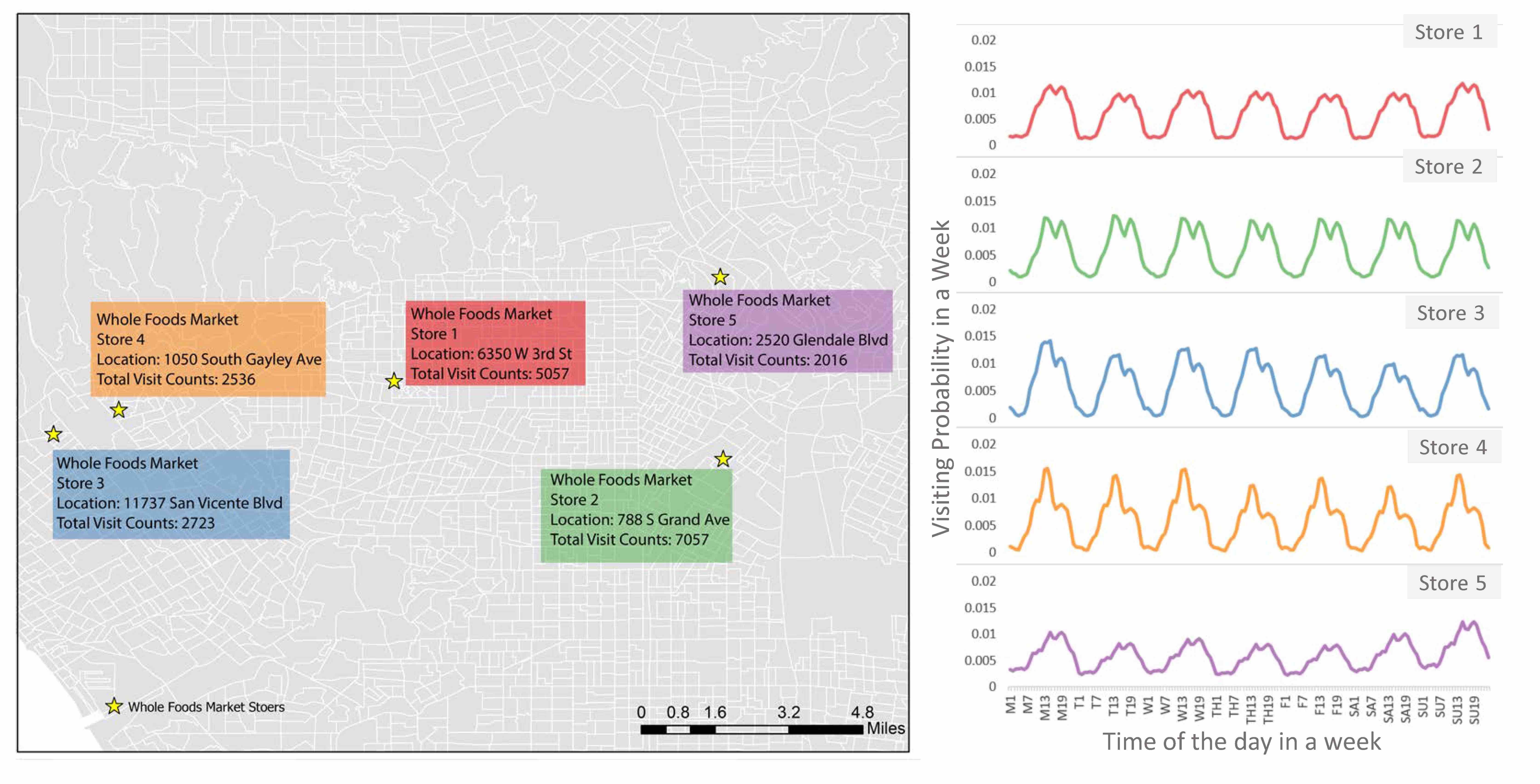}
	\caption{The Whole Foods Markets in Los Angeles with their temporal visit probability plots.}
	\label{fig:hourly_plot}
\end{figure}

Given that people visit different places of interest at different times \citep{mckenzie2015regional,mckenzie2015poi}, we propose the following time-aware dynamic Huff model:

\begin{equation}\label{equation:huff}
P_{ijt} = \frac{\frac{S_{j}^\alpha}{D_{ij}^\beta}}{\sum_{j=1}^{n}\frac{S_{j}^\alpha}{D_{ij}^\beta}}*P_{jt}
\end{equation}
\begin{equation}
P_{jt} = \frac{V_{jt}}{\sum_{t=1}^{m}V_{jt}}
\end{equation}
where $P_{ijt}$ is the probability of a customer $i$ visiting a store $j$ within a temporal window $t$ (e.g., a hour or a day of week); $S_j$ is the attractiveness of the store $j$; $D_{ij}$ is the physical distance between the customer $i$ and the store $j$; $P_{jt}$ is the temporal visit probability for one store $j$ within a temporal window $t$. $V_{jt}$ is the total visit counts for one store $j$ within a particular hour $t$ (in this research) and we sum up the counts over one week as $\sum_{t=1}^{m}V_{jt}$ (i.e., $m$ = 168 hours). As shown in Figure \ref{fig:hourly_plot}, even for the same brand of chain-store (e.g., Whole Foods), the five branch stores in Los Angeles have distinct temporal visit patterns. The parameters $\alpha$ and $\beta$ are used to reflect the effects of attractiveness and the distance on the model.

We also construct another advanced time-aware dynamic Huff (A-Huff) model which estimates the customer visiting probability at each timestamp by comparing all possible visits the customer may have at the same timestamp, which considers the business competition from integrated spatial and temporal aspects. The A-Huff model shares the same parameters with the T-Huff model in equation (\ref{equation:huff}) but with a different formulation as follows. 

\begin{equation}
P_{ijt} = \frac{\frac{S_{j}^\alpha}{D_{ij}^\beta}*P_{jt}}{\sum_{j=1}^{n}\frac{S_{j}^\alpha}{D_{ij}^\beta}*P_{jt}}
\end{equation}
\begin{equation}
P_{jt} = \frac{V_{jt}}{\sum_{t=1}^{m}V_{jt}}
\end{equation}

In addition to the predicted visiting probability $P_{ijt}$ using the A-Huff model, the actual visiting probability $P_{ijt}'$ for this model is calculated using the formula below.
\begin{equation}
P_{ijt}' = \frac{V_{ij}*P_{jt}}{\sum_{j=1}^{n} V_{ij}*P_{jt}}
\end{equation}
where $V_{ij}$ is the observed pairwise visits from the customer $i$ in a specific neighborhood to the store $j$.

\subsection{Parameter calibration using PSO}
Before we use the original Huff and the time-aware dynamic models (T-Huff and A-Huff) to make market share predictions, we need to calibrate the models by adjusting their parameters to make sure that the results approximate or reflect the reality. Previously the two parameters ($\alpha$ and $\beta$) are often decided arbitrarily, which may lead to inaccurate or even erroneous results \citep{huff2003parameter}. A few methods have been used to find a optimized set of $\alpha$ and $\beta$. Many researches used the ordinary-least-squares (OLS) method to estimate the parameters by transforming the Huff model into a logarithm-centering format and estimate the parameters using linear regression \citep{nakanishi1974parameter, huff2008calibrating}. The geographically weighted regression (GWR) was also used to calibrate the Huff model which estimated the parameters for every point inside the study area \citep{suarez2015locating}. Recent research applied optimization algorithms such as the Particle Swarm Optimization (PSO) technique to find optimal or near-optimal solution of parameters that fit the observation data more accurately \citep{suhara2019validating}.

In this research, we used the PSO technique for calibrating both the Huff and the T-Huff models' parameters, which was introduced by \cite{eberhart1995new}, inspired by the foraging behavior of bird flocking. As a widely used optimization method, PSO makes few or no assumptions (e.g., linearity) about the problem being optimized, so it is appropriate for our problem. Also, we are able to design the objective function based on different needs. In our case, we selected the correlation between the predicted store visit probability and the actual visit probability as the objective function. Compared with the traditional OLS approach, the PSO technique allows more freedom at the optimization design stage and is efficient to find the solutions from a very large space of candidate solutions, which means that we can try a great number of $\alpha$ and $\beta$ values and observe the trend of convergence through the optimization process. Therefore, we selected the PSO as the optimization method to find the optimal $\alpha$ and $\beta$ in this work.
To initialize the optimization, a few particles are generated, and each particle represents a pair of potential $\alpha$ and $\beta$. The particles will change their positions (which is the value of $\alpha$ and $\beta$) based on its previous best location and the global best position \citep{xiao2013reconstructing, kennedy2010particle}. The particles should then gradually cluster in the area of the optimal solution and return a optimized result. Here the performance of every particle is determined by a pre-defined objective function. The goal of the optimization process is to find the best combination of parameters that maximize the objective function. The objective function in this study is the Pearson correlation between the estimated probability and the actual probability of pairwise visits from a particular neighborhood to a store. We calibrate the parameters for each specific brand of stores using large-scale anonymous mobile phone location tracking data (in the following section \ref{sec:data}) respectively in order to find the models that can best reflect the particular store visit patterns.

\section{Data and study area} \label{sec:data}
\begin{figure}[H]
\small
\centering
\begin{subfigure}[b]{0.8\textwidth}
  \centering
  \includegraphics[width=\textwidth]{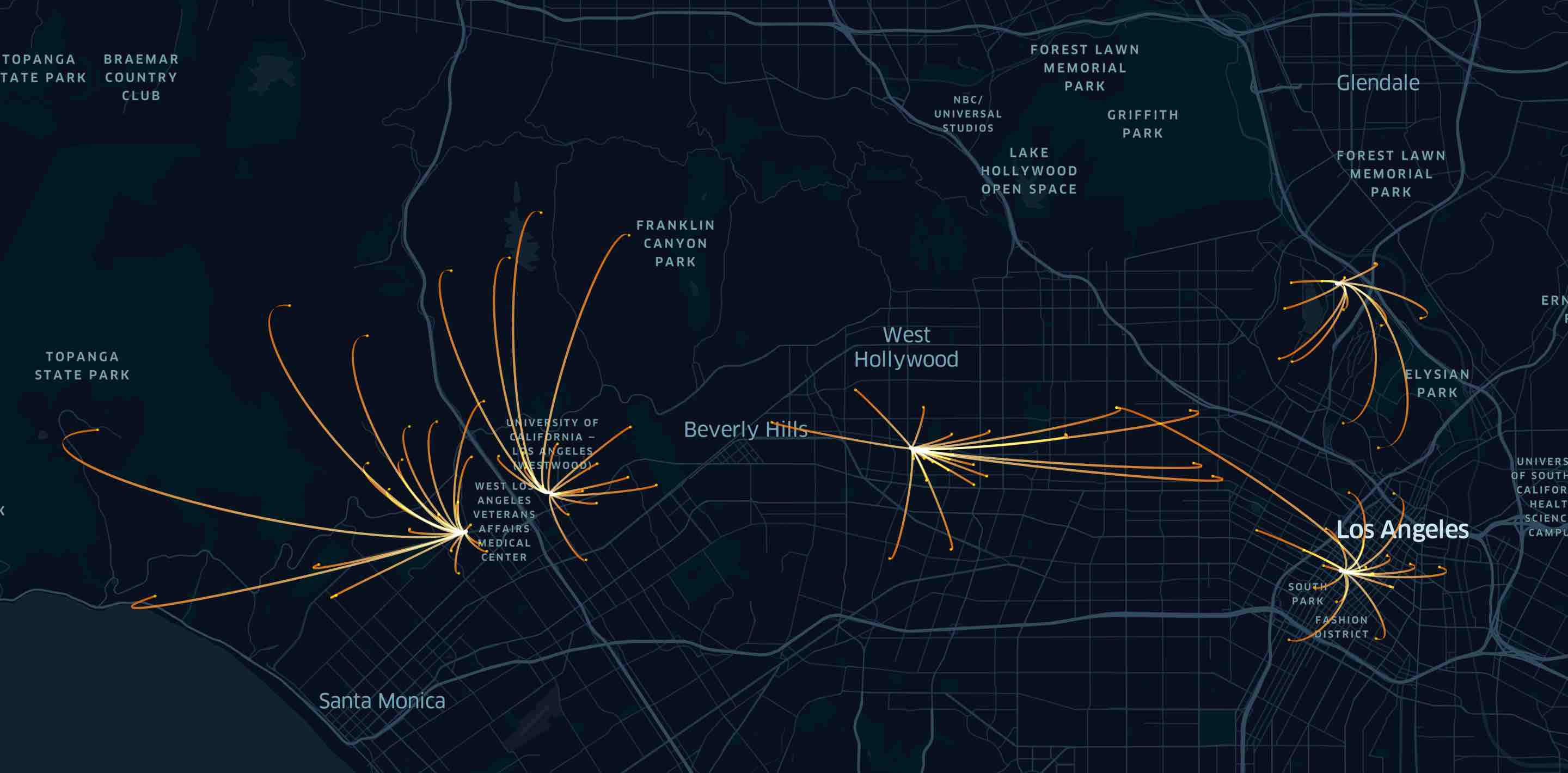}
  \caption{}
  \label{fig:map_arc_wholefoods}
\end{subfigure}
     \hfill
\begin{subfigure}{.8\textwidth}
  \centering
  \includegraphics[width=\textwidth]{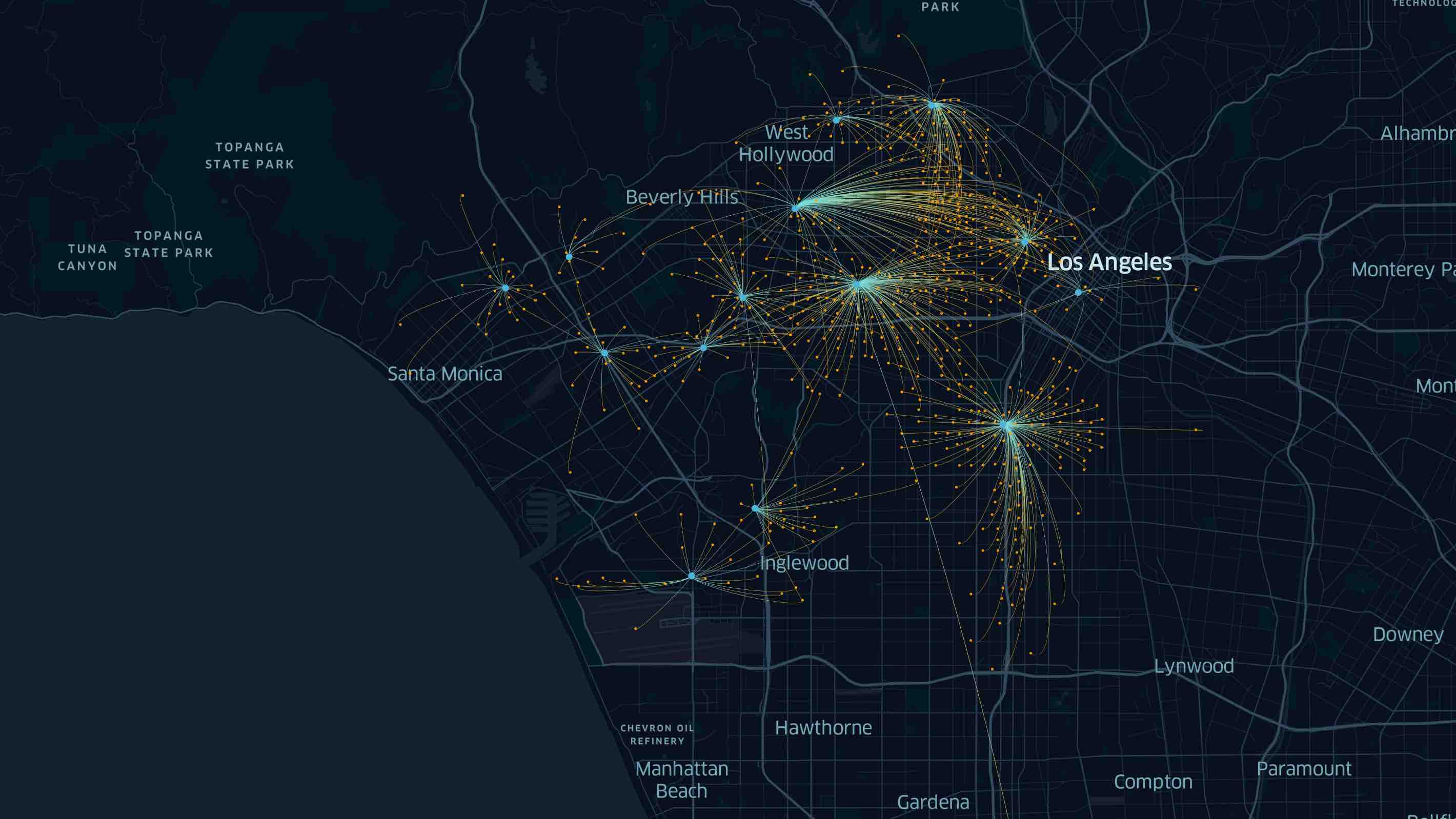}
    \caption{}
 \label{fig:map_arc_ross}
\end{subfigure}
	\caption{The spatial distributions of CBGs that have visit flows to (a) five Whole Foods Markets and (b) fourteen Ross Stores in Los Angeles (Note: the number of stores for each brand only reflects the data we have; the geovisualization is created using the kepler.gl tool).}
	\label{fig:map_arc}
\end{figure}

We collected over 3.6 million points of interest (POIs) with visit patterns in the U.S. from the SafeGraph business venue database\footnote{\url{https://www.safegraph.com}}. The POIs are first classified based on the North American Industry Classification System (NAICS) 6-digit sector codes. Among them, we selected two categories of interest:``445110-Supermarkets and Grocery Stores" and``452210-Department Stores". There are over 20,000 POIs selected in the top ten most populated U.S. cities (New York, Los Angeles, Chicago, Houston, Phoenix, Philadelphia, San Antonio, San Diego, Dallas, and San Jose). In addition to the spatial distribution of the POIs, we also retrieved the fine-resolution visit patterns of all those POIs from the aforementioned SafeGraph database which covers dynamic human mobility patterns of millions of anonymous smart phone users. The SafeGraph’s data sampling correlated highly with the U.S. Census populations\footnote{\url{https://www.safegraph.com/blog/what-about-bias-in-the-safegraph-dataset}}. These mobile location data consist of “pings” identifying the coordinates of a smartphone at a moment in time. To enhance privacy, SafeGraph excludes census block group (CBG) information if fewer than five devices visited a place in a month from a given CBG. For each POI, the records of aggregated visitor patterns illustrate the number of unique visitors and the number of total visits to each venue during the specified time window (i.e., October to December 2018 in our dataset), which could reflect the attractiveness of each venue. For example, Figure \ref{fig:map_arc} shows the spatial distributions of CBGs that have visit flows to the five Whole Foods Markets and the fourteen Ross Stores in Los Angeles. Furthermore, we also computed the average of hourly visit probability for each POI over 168 hours (24 hours * 7 days of a week) to show the dynamic visit patterns. For future studies, the hourly visit frequency can also be estimated from other resources, such as the shopper's loyalty card data or the popular times collected by Google Maps or Yelp for business locations.  The corresponding demographic and socioeconomic attribute data of all CBGs were collected from the American Community Survey (ACS).

\section{Results} \label{sec:results}
\subsection{Visit distance distributions} \label{sec:dist_distribution}
\begin{figure}[H] 
	\centering
	\includegraphics[width=1.0\textwidth]{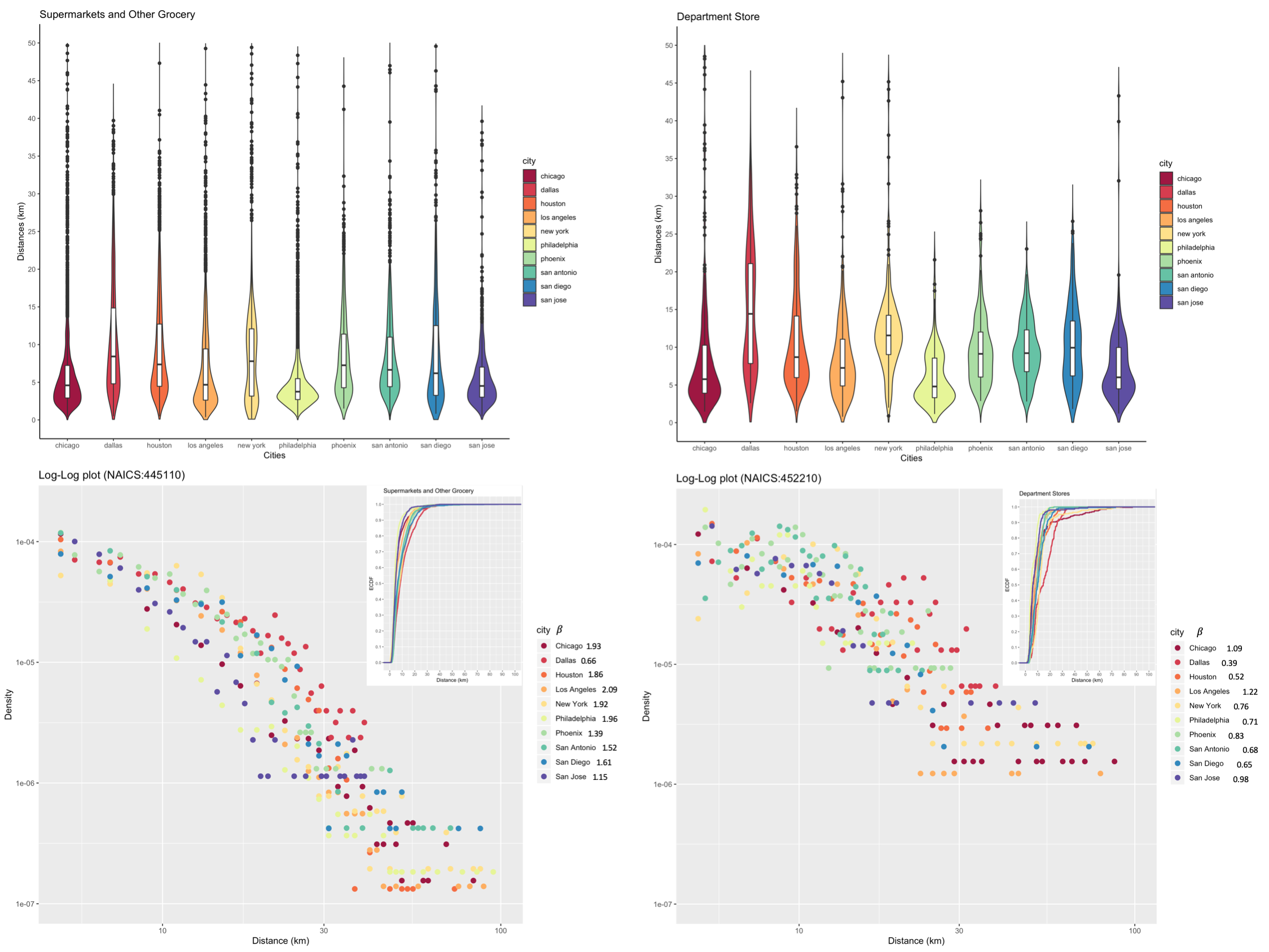}
	\caption{The probability density distribution, empirical cumulative distribution, and log-log plots of visitors' distance from home to supermarkets and grocery stores (NACIS: 445110) and to department stores (NACIS: 452210) in the top 10 most populated cities in US.}
	\label{fig:distancefromhome}
\end{figure}

We first analyzed the distribution of the median distance that visitors traveled from home to all the stores given a specific NACIS category. The probability density distributions of visit distances across cities showed a variety of heavy tailed distributions. The mean of the median distance (great circle distance) from visitors' home to supermarkets and grocery stores (NACIS: 445110) across these cities is about 7.8 km. However, the median distance distribution does vary over different cities (as shown in Figure\ref{fig:distancefromhome}). Most people in Philadelphia, San Jose, Chicago, and Los Angeles traveled relatively shorter distances, with the median of 3.8 km, 4.5 km, 4.6 km, and 4.7 km respectively, than people in other big metropolitan areas in US such as Dallas and New York with the largest median distance of 8.4 km and 7.8 km respectively. As expected, the mean of the median distance from visitors' home to department stores (NACIS: 452210) across these cities is about 10.3 km and larger than that to supermarkets and grocery stores.

In addition, the distance decay phenomenon exists in the visit median distance density distribution across all cities (as shown in the log-log plots in Figure \ref{fig:distancefromhome}). The visit probability decreased significantly after about 10 km, which offers insights into location business decision makings. And different cities have varying decay exponents $\beta$ \citep{gao2013discovering}, which may link to their urban morphology (e.g., size and shape) \citep{kang2012intra}. The distance decay slopes for supermarkets \& grocery stores are steeper than that of department stores in all cities, which demonstrate that there are much fewer long-distance travels for supermarkets \& grocery store visits compared with that for department stores. 

\subsection{Huff models calibration for top brands}
\subsubsection{Parameter calibration and comparison}
Given the variability of store visits in chain-store brands and local brands in our exploratory analysis, we didn't calibrate the models for all brands in each POI category. Instead, we only designed comparative experiments for top three chain-store brands with the most stores across the ten most populated cities in our dataset. Take the Whole Foods Markets in Los Angeles as an example, the attractiveness of each Whole Foods store is estimated using the total visit count over three months in the Safegraph dataset. Figure \ref{fig:map_arc_wholefoods} shows the flow map from each CBG to the five Whole Foods Markets in the Los Angeles area. It is clear that people in each CBG has a particular store visit preference, and the visited store is usually within certain spatial proximity to that CBG. The Whole Foods Markets are chain stores that usually have similar product layouts and sizes. Therefore, the major factor affecting the visits of customers is usually the distance from the customer to the store. There are also some other factors. For example, for the two Whole Foods Markets on the left part of Figure \ref{fig:map_arc_wholefoods}, we can see a clear delineation of visiting CBGs to the two stores separated by the highway. Even though these two Whole Foods Markets are located closely to each other, they have very distinct visitors due to such infrastructure barrier in that area. Other demographic and socioeconomic factors influencing the store visits will be further discussed in section \ref{sec:regression}.

The model parameter calibration is conducted for each brand of stores respectively in order to find the best set of $\alpha$ and $\beta$ that can reflect the effects of attractiveness and distance on the particular brand using observed store visit data. 
A set of values for $\alpha$ and $\beta$ is first determined in order to identify a smaller data range for optimization. The result of the correlation for the selected $\alpha$ and $\beta$ for the original Huff model is shown in  Table \ref{tab:original_cor_result}. In general, the model produces very good results with all Pearson's correlation coefficients larger than 0.6. A higher correlation is obtained with $\alpha$ between 0 to 1 and $\beta$ between 0 to 2 approximately. Therefore, the bounds for $\alpha$ and $\beta$ in the PSO optimization are set to be from 0 to 2. The optimization is repeated 10 times with 10 particles and is implemented using a open-source library ``Pyswarms" in Python. The highest correlation obtained from the optimization is 0.864 when $\alpha = $ 0.717 and $\beta = $ 0.805. The $\alpha$ and $\beta$ values are then fed into the Huff model to estimate the store visit probability.

\begin{table} [!h]
\centering
	\caption{The model parameter calibration results with Pearson's correlation for Whole Foods using the original Huff model.}
\begin{tabular}{|c|c|c|c|c|c|}
\hline
  \diagbox[height=5ex, width=4em]{\raisebox{0.4\height}{\enspace $\alpha$}}{ \raisebox{-0.5\height}{\  $\beta$}} 
 &  0.1 & 0.5 & 1 & 2 & 5 \\ \hline
  0.1  & 0.807 & 0.844 &  0.845 & 0.817 & 0.769 \\\hline
  0.5 & 0.808 & 0.854 &  0.858 & 0.825 & 0.774 \\\hline
  1  & 0.791 & 0.846 &  0.862 & 0.828 & 0.778 \\\hline
  2  & 0.747 & 0.797 &  0.834 & 0.822 & 0.776 \\\hline
  5 & 0.683 & 0.709 &  0.740 & 0.773 & 0.752\\
\hline
\end{tabular}
\label{tab:original_cor_result}
\end{table}

Also, table \ref{tab:THuff_cor_result} shows the Pearson correlation result with the same selected $\alpha$, $\beta$ values using the T-Huff model. In general, the result from the T-Huff model has higher correlations for all selected $\alpha$ and $\beta$ than the original Huff model, which reflects that the T-Huff model might provide a more accurate estimation of the dynamic visit probability in most of the cases. The highest correlation obtained from the optimization procedure is 0.890 with $\alpha$ = 0.787 and $\beta$ = 0.765. 

\begin{table} [!h]
\centering
	\caption{The model parameter calibration results with Pearson's correlation for Whole Foods using the T-Huff model.}
\begin{tabular}{|c|c|c|c|c|c|}
\hline
  \diagbox[height=5ex, width=4em]{\raisebox{0.4\height}{\enspace $\alpha$}}{ \raisebox{-0.5\height}{\  $\beta$}} 
 &  0.1 & 0.5 & 1 & 2 & 5 \\ \hline
  0.1  & 0.847 & 0.874 &  0.873 & 0.844 & 0.791 \\\hline
  0.5 & 0.848 & 0.882 &  0.884 & 0.852 & 0.796 \\\hline
  1  & 0.835 & 0.877 &  0.888 & 0.855 & 0.801 \\\hline
  2  & 0.789 & 0.832 &  0.861 & 0.847 & 0.799 \\\hline
  5 & 0.694 & 0.716 &  0.744 & 0.775 & 0.761\\
\hline
\end{tabular}
\label{tab:THuff_cor_result}
\end{table}

In addition to the original static Huff model and the dynamic Huff models (T-Huff and A-Huff), another model named as the M-Huff model is constructed for comparison. The synthetic M-Huff model assumes that the hourly visit probability for one CBG to one store is distributed evenly over the 168 hours in one week (using the mean visit probability) and therefore the model assigns the visit probability equally to each time window (every hour in this study). The correlation is then calculated between this equally-distributed visit probability and the actual hourly visit probability from the SafeGraph dataset. Table \ref{tab:SHuff_cor_result} shows the correlation result for selected $\alpha$, $\beta$ from the M-Huff model. The highest correlation from the optimization is 0.662 with $\alpha$ = 0.723 and $\beta$ = 0.806. It is clear that the correlations drop dramatically compared with the results of the original Huff model and the T-Huff model, which means that the assumed equally-distributed hourly visit probability can not make a good representation of the actual dynamic visit patterns. In other words, the store visit patterns do have temporal variation and it is necessary to consider such variation in market-share models.

\begin{table} [!h]
\centering
	\caption{The model parameter calibration results with Pearson's correlation for Whole Foods using the M-Huff model.}
\begin{tabular}{|c|c|c|c|c|c|}
\hline
  \diagbox[height=5ex, width=4em]{\raisebox{0.4\height}{\enspace $\alpha$}}{ \raisebox{-0.5\height}{\  $\beta$}} 
 &  0.1 & 0.5 & 1 & 2 & 5 \\ \hline
  0.1  & 0.618 & 0.646 &  0.647 & 0.626 & 0.589 \\\hline
  0.5 & 0.619 & 0.654 &  0.657 & 0.632 & 0.593 \\\hline
  1  & 0.606 & 0.648 &  0.660 & 0.634 & 0.596 \\\hline
  2  & 0.572 & 0.610 &  0.639 & 0.629 & 0.594 \\\hline
  5 & 0.523 & 0.543 &  0.566 & 0.592 & 0.576\\
\hline
\end{tabular}
\label{tab:SHuff_cor_result}
\end{table}

\subsubsection{Visit spatial pattern comparison}

\begin{figure}[H]
\small
\centering
\begin{subfigure}[b]{0.75\textwidth}
  \centering
  \includegraphics[width=\textwidth]{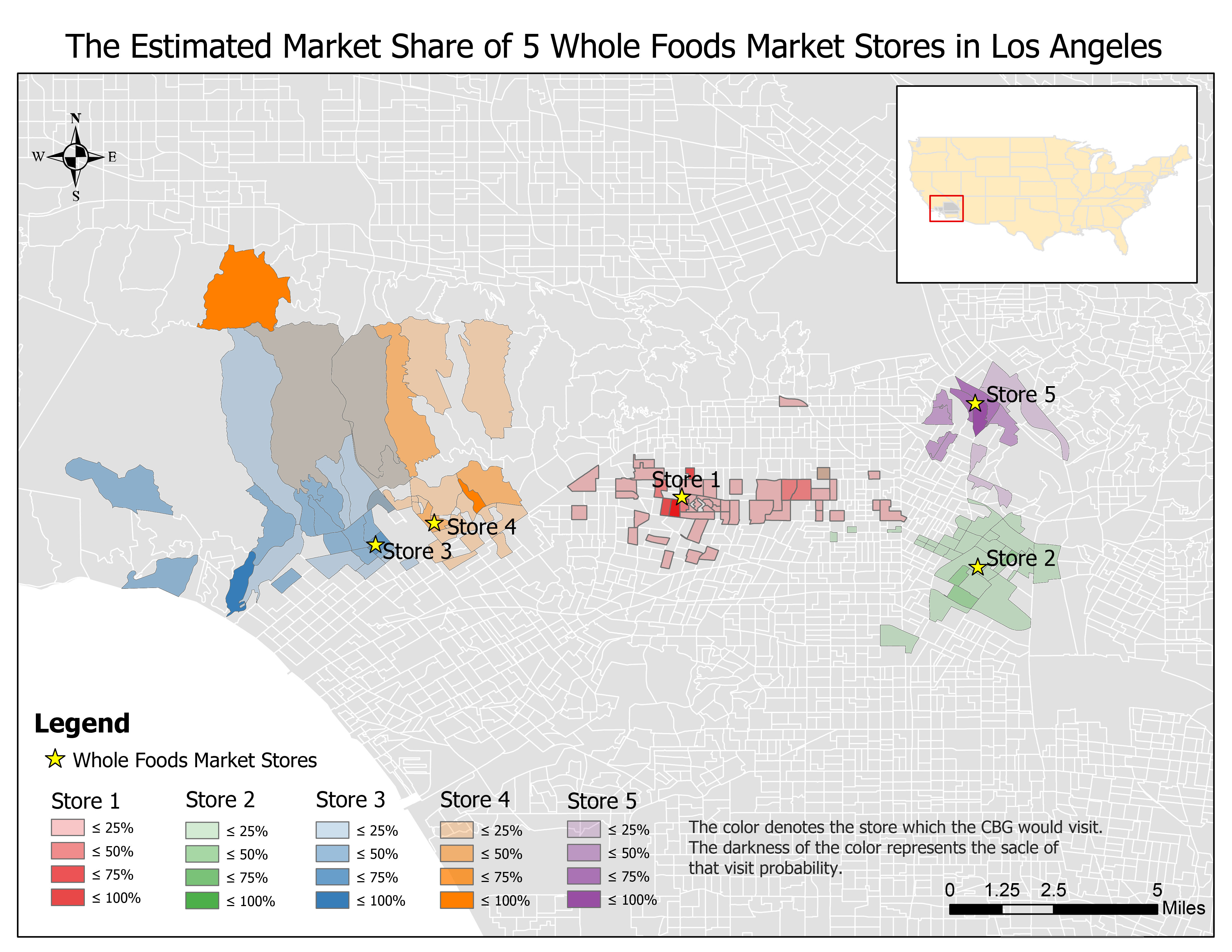}
  \caption{}
  \label{fig:map_estimated}
\end{subfigure}
     \hfill
\begin{subfigure}{0.75\textwidth}
  \centering
  \includegraphics[width=\textwidth]{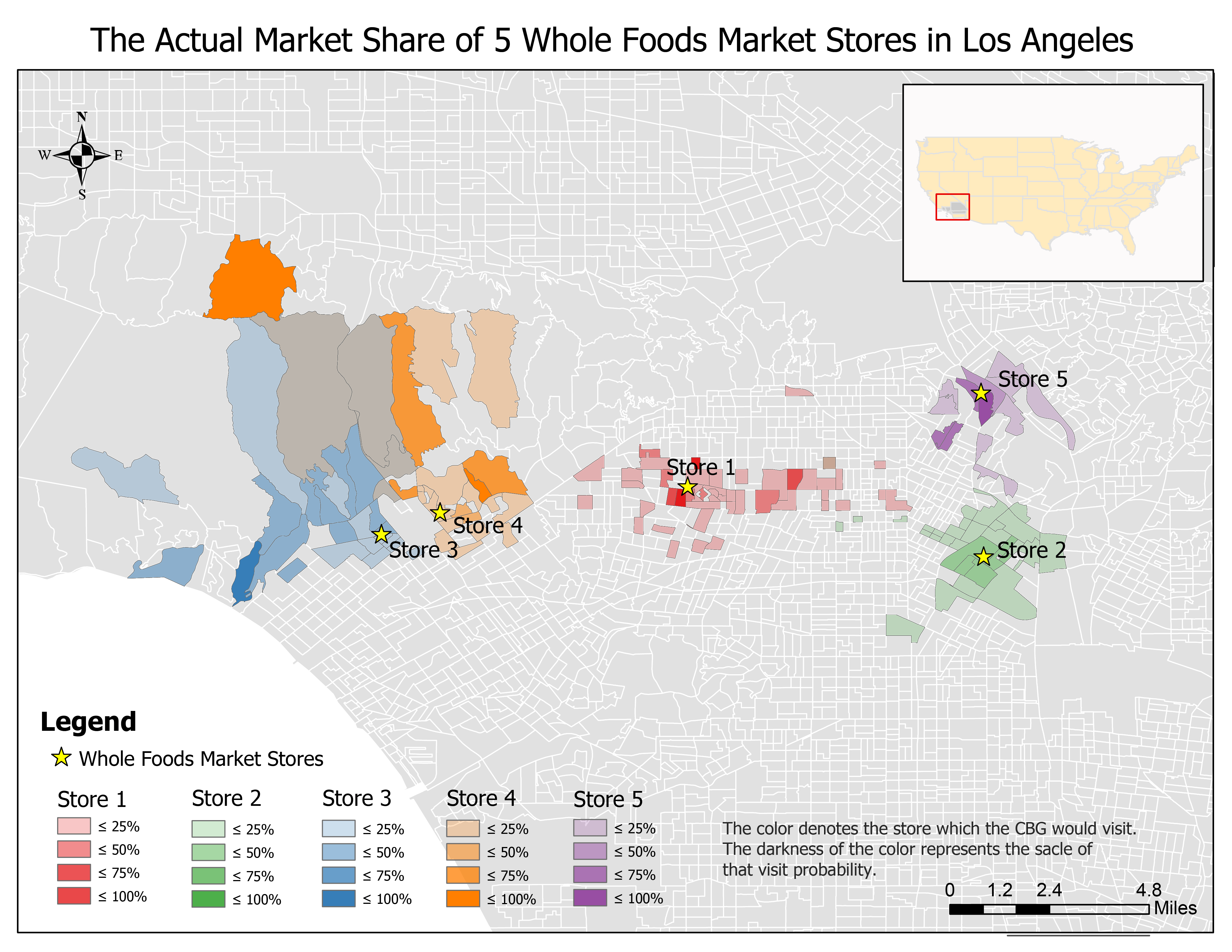}
    \caption{}
 \label{fig:map_actual}
\end{subfigure}
\caption{The estimated market share using the original Huff model and the actual market share derived from the SafeGraph visit database.}
\label{fig:map_huff_wholefoods}
\end{figure}

Figure \ref{fig:map_huff_wholefoods} shows two maps of the estimated market share from the original Huff model and the actual market share generated from the SafeGraph POI visit dataset. Here the market share means the probability that people from a CBG will visit that particular store. For every CBG, it has a corresponding visit probability for each store, and the color hue of each CBG represents the store that people from this CBG would visit. The saturation of the color indicates the magnitude of the probability. By comparing the two maps, we find that the spatial distributions of trade areas are very similar (with high correlation of store visit probabilities). It means that the estimated result from the original Huff model can project the total visit probability with high accuracy. The result also supports our above statement that the large portion of visitors of each Whole Foods Market are usually within close proximity of that store. People may be reluctant to go to another Whole Foods Market that is far away from them. This is one characteristic of the chain stores that the location of a store is very important to the performance of that store. As the chain stores may not be very different from each other regarding their products, the spatial proximity between the store and the customer becomes a primary factor affecting people's choice. 

\begin{figure}[H]
\centering
\begin{subfigure}{.5\textwidth}
  \centering
  \includegraphics[width=\textwidth]{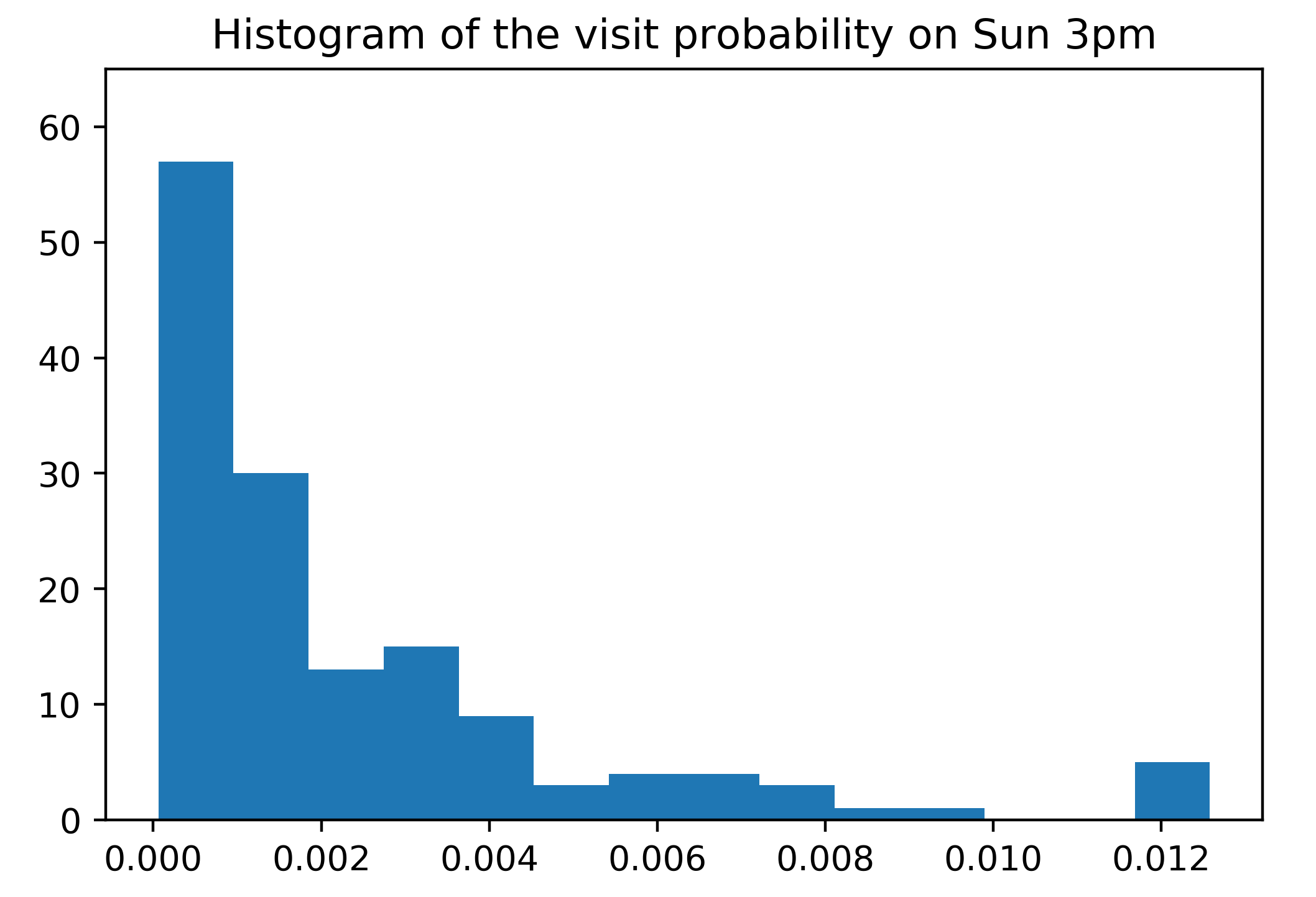}
  \caption{}
  \label{fig:sun15_hist}
\end{subfigure}%
\begin{subfigure}{.5\textwidth}
  \centering
  \includegraphics[width=\textwidth]{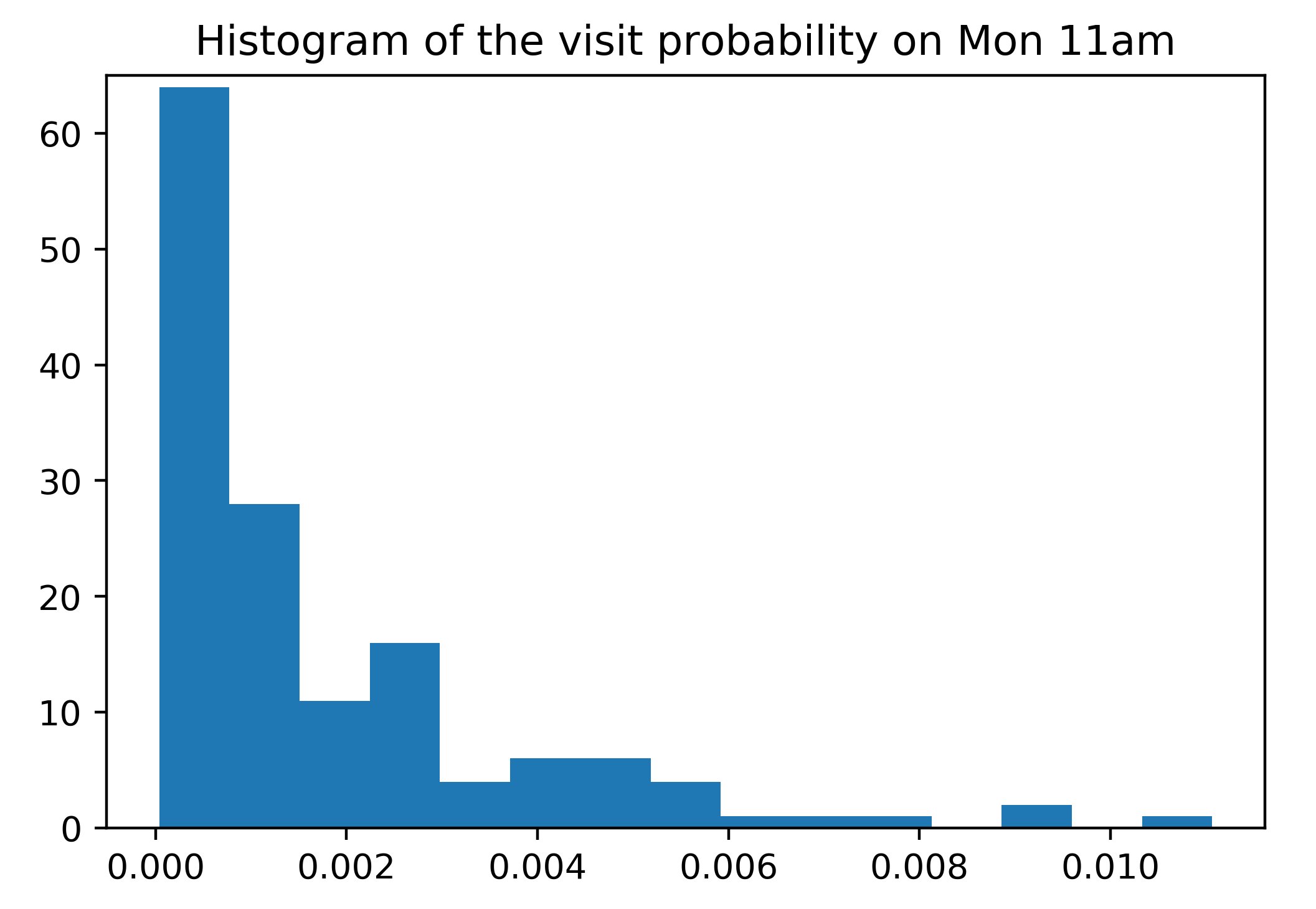}
  \caption{}
  \label{fig:mon11_hist}
\end{subfigure}
\caption{The histograms of the visit probability on (a) Sunday 3:00pm-3:59pm and (b) Monday 11:00am-11:59am.}
\label{fig:hist_visit_prob}
\end{figure}

\begin{figure}[H]
\small
\centering
\begin{subfigure}[b]{0.7\textwidth}
  \centering
  \includegraphics[width=\textwidth]{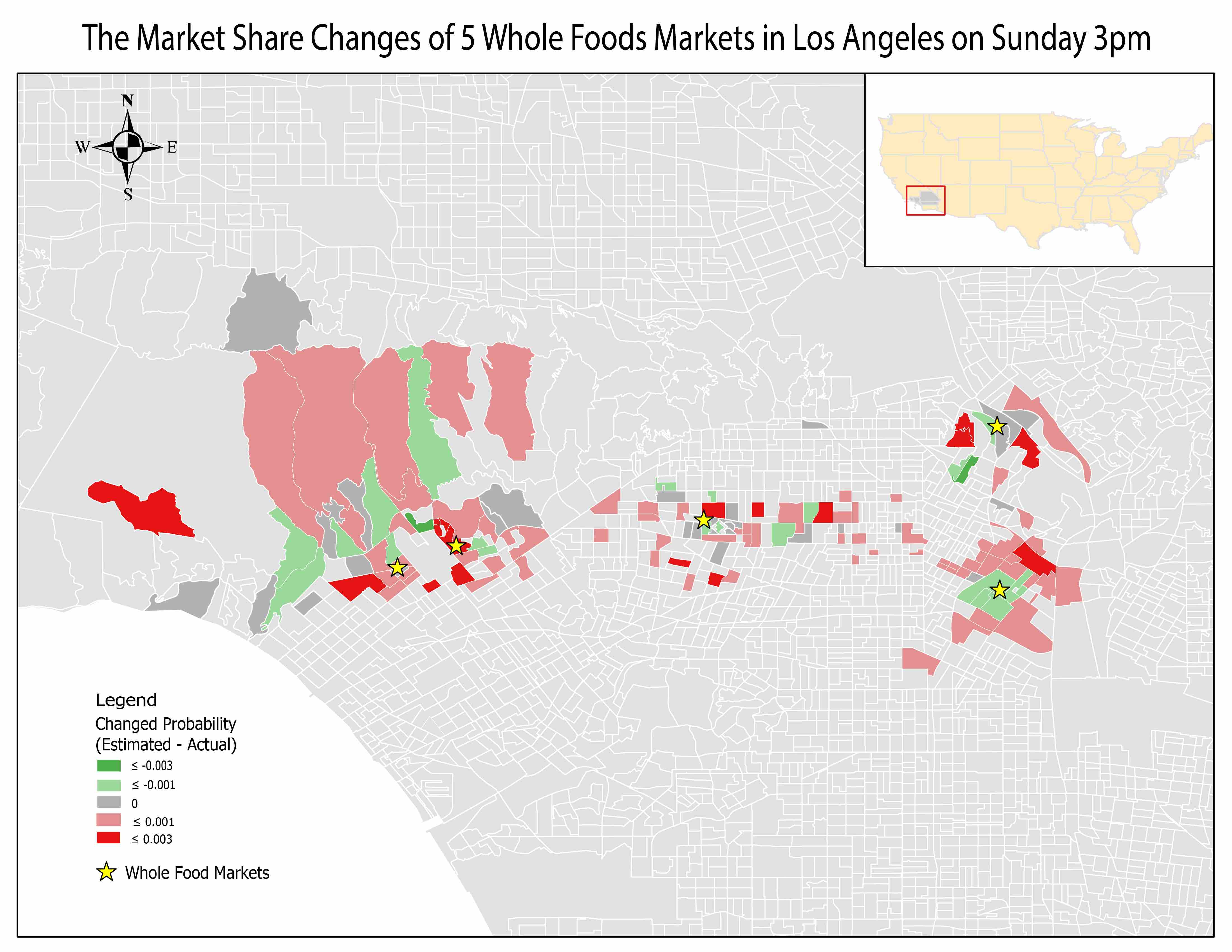}
  \caption{}
  \label{fig:sun15_hist}
\end{subfigure}

\begin{subfigure}[b]{0.7\textwidth}
  \centering
  \includegraphics[width=\textwidth]{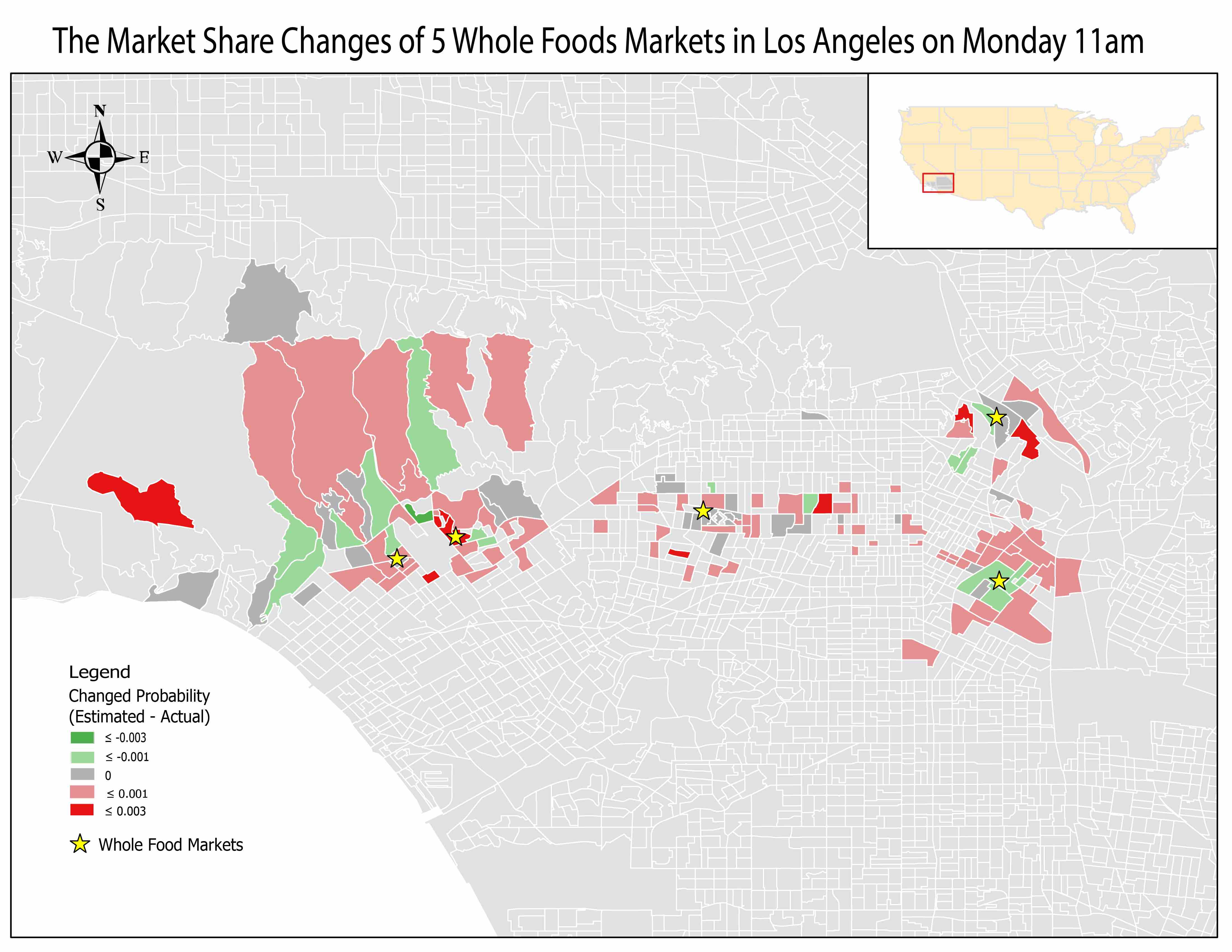}
  \caption{}
  \label{fig:sun15_hist}
\end{subfigure}
\caption{The maps of the visit probability changes between the estimated market share using the T-Huff model and the actual market share derived from the SafeGraph visit database on (a) Sunday 3:00pm-3:59pm and (b) Monday 11:00am-11:59am.}
\label{fig:wholefoods_thuff_plots}
\end{figure}

 Figure \ref{fig:hist_visit_prob} shows the histograms of hourly visit probability on Sunday 3:00pm-3:59pm and Monday 11:00am-11:59am. Figure \ref{fig:wholefoods_thuff_plots} maps the difference between the estimated and the actual market share of the Five Whole Foods on two different time windows obtained from the dynamic Huff model. Here we pick two different hours (Sunday 3:00pm-3:59pm and Monday 11:00am-11:59am) to compare how the POI visit probability may differ in different time of a day and different day of a week \citep{mckenzie2015poi}. The data classification intervals for the visit probability mapping are determined by geometrical intervals as the probability distribution for all CBGs to all Whole Foods stores in the two hours both follow a right-skewed distribution.

From the T-Huff model, as the visit probability is assigned to a specific hourly window, it has a much smaller range compared with that of the original Huff model. Therefore, the ranges of the probability differences are also smaller, usually between -0.003 to 0.003 from the maps in Figure \ref{fig:wholefoods_thuff_plots}. Also, we can see that most of the prediction errors are between the ranges of -0.001 to 0.001. The prediction for Monday 11am has a better accuracy than the one for Sunday 3pm as we can see that there are less dark red or dark green areas on the map for Monday 11am. One reason could be there is larger variability of visits on Sunday 3pm.

\subsubsection{Brands comparison and regional variability}
The same process of model parameter calibration using PSO for three brands (Whole Foods, Trader Joe's and Ross Stores) are conducted for the ten U.S. cities. Three types of comparisons are examined: (1) comparing the performance of four models; (2) comparing how the models perform differently for the three brands; (3) and discovering whether there exists regional variability among the same type of stores across different cities. Table \ref{tab:store_number} shows the number of stores for the three brands in each city.

\begin{table} [!h]
\small
\centering
	\caption{The number of stores for the three brands in ten cities. (Note: the number only reflects the data we have.)}
\begin{tabular}{|c|c|c|c|}
\hline
 & Whole Foods & Trader Joe's & Ross Stores \\\hline
  Los Angeles  & 5 & 11 & 14 \\\hline
  Houston & 7 & 3 & 24  \\\hline
  Chicago & 10 & 5 & 12  \\\hline
  Philadelphia & 2 & 1 & 8  \\\hline
  New York & 8 & 5 & 0\\\hline
  San Antonio & 1 & 2 & 15\\\hline
  Dallas & 4 & 4 &7\\\hline
  San Diego & 1 & 6 & 7\\\hline
  San Jose & 2 & 4 & 6\\\hline
  Phoenix & 3 & 1 & 15 \\
\hline
\end{tabular}
\label{tab:store_number}
\end{table}

Table \ref{tab:cor_result} shows the highest correlation coefficients from the PSO for the four Huff models and three brands in the ten cities. Tables \ref{tab:parameter_result} and \ref{tab:parameter_result_alpha} show the corresponding $\alpha$ and $\beta$ values for each optimal solution.  

By looking at each row, we compare the performance of four models. The optimal correlations are generally high for the original Huff model, the T-Huff model and the A-Huff model across all stores and cities. But the correlation from the M-Huff model is always much lower than that of the other three models, which indicates that the temporal variation cannot be ignored or simply considered as equally distributed. The T-Huff model and the A-Huff model have slightly higher correlation than the traditional Huff model, which can show that the temporal variation is important and can help improve the estimation accuracy. The result of the T-Huff model is the highest among the four models for each single brand and each city in most of the cases, which shows the way of adding the temporal visiting information in this model yields the best performance in our study. By comparing the parameters in Tables \ref{tab:parameter_result} and \ref{tab:parameter_result_alpha}, the optimal $\alpha$ and $\beta$ remain similar for each brand in each city among four models. This indicates that for each particular type of POIs in each city, the optimization process is able to find consistent parameters among four models that reflect the impacts of attractiveness and distance specifically for each brand in that city.

\begin{table} [!h]
\small
\centering
	\caption{The optimized correlation for three brands.}
\resizebox{1.35\textwidth}{!}{
\begin{tabular}{|c|c|c|c|c|c|c|c|c|c|c|c|c|}
\hline
 & \multicolumn{4}{c|}{Whole Foods} &  \multicolumn{4}{c|}{Trader Joe's} & \multicolumn{4}{c|}{Ross Stores}\\ \hline
    & Huff & M-Huff & T-Huff & A-Huff & Huff & M-Huff & T-Huff & A-Huff& Huff & M-Huff & T-Huff & A-Huff\\\hline
  Los Angeles  & 0.864 & 0.662 &  0.890 & 0.878 & 0.875 & 0.588 & 0.910 & 0.900 & 0.854&0.664 & 0.881 & 0.863\\\hline
  Houston & 0.827 & 0.567 &  0.874 &0.838 & 0.682 & 0.391 & 0.827 &0.776 & 0.821 & 0.544 & 0.864 &0.845 \\\hline
  Chicago  & 0.869 & 0.580 &  0.904 & 0.892 & 0.892 & 0.578 & 0.919 &0.913 & 0.933 & 0.670 & 0.946 &0.940 \\\hline
  Philadelphia  & 0.869 & 0.612 &  0.899 &0.892 & 0.956 & 0.770 & 0.962 & 0.959 & 0.892 & 0.637 & 0.917 &0.904 \\\hline
  New York & 0.949 & 0.602 &  0.968 &0.955 & 0.847 & 0.567 & 0.902 &0.863 & NA & NA & NA &NA\\\hline
  San Antonio & 0.888 & 0.434 &  0.931 &0.923 & 0.644 & 0.406 & 0.748 &0.721 & 0.935 & 0.650 & 0.942 & 0.942\\\hline
  Dallas & 0.901 & 0.573 &  0.932 &0.908 & 0.948 & 0.603 & 0.963 &0.960 & 0.953 & 0.610 & 0.965 &0.960\\\hline
  San Diego & 0.825 & 0.614 &  0.853 &0.833 & 0.919 & 0.604 & 0.938 &0.929 & 0.917 & 0.648 & 0.928 &0.925\\\hline
  San Jose & 0.959 & 0.687 &  0.964 & 0.964 & 0.927 & 0.563 & 0.952 &0.947 & 0.903 & 0.582 & 0.930 &0.924\\\hline
  Phoenix & 0.966 & 0.589 & 0.979 &0.967 &  0.959 & 0.571 & 0.970 & 0.970 & 0.900 & 0.615 &0.924 & 0.910\\\hline
  Average & 0.8917 & 0.592 & 0.919 & 0.905 & 0.865 & 0.564 & 0.909 & 0.894 & 0.901 & 0.624 & 0.922 & 0.9126 \\
\hline
\end{tabular}
}
\\`NA': no data available 
\label{tab:cor_result}
\end{table}

\begin{table} [!h]
\setlength{\tabcolsep}{4pt}
\centering
	\caption{The optimized Huff model coefficients $\beta$ for three brands.}
\resizebox{1.35\textwidth}{!}{
\begin{tabular}{|c|c|c|c|c|c|c|c|c|c|c|c|c|}
\hline
             & \multicolumn{4}{c|}{Whole Foods} & \multicolumn{4}{c|}{Trader Joe's} & \multicolumn{4}{c|}{Ross Stores} \\ \hline
             & Huff  & M-Huff & T-Huff & A-Huff & Huff  & M-Huff  & T-Huff & A-Huff & Huff  & M-Huff & T-Huff & A-Huff \\ \hline
LosAngeles   & 0.8   & 0.81   & 0.76   & 0.85   & 0.56  & 0.6     & 0.51   & 0.59   & 0.93  & 1.1    & 1.01   & 0.93   \\ \hline
Houston      & 0.91  & 1.01   & 0.9    & 0.92   & 0.74  & 0.74    & 0.7    & 0.97   & 0.67  & 0.66   & 0.63   & 0.63   \\ \hline
Chicago      & 0.6   & 0.57   & 0.66   & 0.56   & 0.64  & 0.58    & 0.68   & 0.54   & 0.73  & 0.57   & 0.65   & 0.68   \\ \hline
Philadelphia & 0.64  & 0.54   & 0.55   & 0.56   & 0.59  & 0.44    & 0.54   & 0.57   & 0.46  & 0.41   & 0.68   & 0.62   \\ \hline
NewYork      & 0.44  & 0.44   & 0.44   & 0.49   & 0.06  & 0.06    & 0.18   & 0.13   &    NA   &    NA    &     NA   &     NA   \\ \hline
SanAntonio   & 0.76  & 0.7    & 0.77   & 0.82   & 0.72  & 0.76    & 0.55   & 0.67   & 0.78  & 0.84   & 0.82   & 0.84   \\ \hline
Dallas       & 0.93  & 0.93   & 0.99   & 1.01   & 0.88  & 0.83    & 0.72   & 0.77   & 0.52  & 0.55   & 0.61   & 0.57   \\ \hline
SanDiego     & 0.04  & 0.03   & 0.03   & 0.14   & 0.93  & 0.84    & 0.93   & 0.94   & 0.43  & 0.44   & 0.49   & 0.45   \\ \hline
SanJose      & 0.84  & 0.95   & 0.84   & 0.78   & 0.98  & 0.85    & 0.82   & 0.72   & 0.78  & 0.8    & 0.78   & 0.75   \\ \hline
Phoenix      & 1.58  & 1.83   & 1.63   & 1.83   & 1.36  & 1.23    & 1.05   & 0.97   & 0.7   & 0.65   & 0.69   & 0.65   \\ \hline
\end{tabular}
}
\\`NA': no data available 
\label{tab:parameter_result}
\end{table}

\begin{table} [!h]
\setlength{\tabcolsep}{4pt}
\centering
	\caption{The optimized Huff model coefficients $\alpha$ for three brands.}
\resizebox{1.35\textwidth}{!}{
\begin{tabular}{|c|c|c|c|c|c|c|c|c|c|c|c|c|}
\hline
             & \multicolumn{4}{c|}{Whole Foods} & \multicolumn{4}{c|}{Trader Joe's} & \multicolumn{4}{c|}{Ross Store's} \\ \hline
             & Huff  & M-Huff & T-Huff & A-Huff & Huff  & M-Huff  & T-Huff & A-Huff & Huff  & M-Huff  & T-Huff & A-Huff \\ \hline
LosAngeles   & 0.72  & 0.72   & 0.79   & 0.69   & 0.44  & 0.44    & 0.45   & 0.54   & 0.59  & 0.77    & 0.73   & 0.62   \\ \hline
Houston      & 0.82  & 0.85   & 0.94   & 0.87   & 0.98  & 0.95    & 0.93   & 0.99   & 0.62  & 0.61    & 0.52   & 0.48   \\ \hline
Chicago      & 0.48  & 0.49   & 0.55   & 0.46   & 0.46  & 0.44    & 0.43   & 0.34   & 0.39  & 0.37    & 0.38   & 0.40   \\ \hline
Philadelphia & 0.88  & 0.98   & 0.84   & 0.83   & 0.36  & 0.35    & 0.31   & 0.37   & 0.6   & 0.54    & 0.8    & 0.74   \\ \hline
NewYork      & 0.39  & 0.3    & 0.28   & 0.33   & 0.26  & 0.12    & 0.2    & 0.17   & NA    & NA      & NA     & NA     \\ \hline
SanAntonio   & 0.78  & 0.89   & 0.72   & 0.71   & 0.83  & 0.9     & 0.68   & 0.69   & 0.58  & 0.6     & 0.66   & 0.67   \\ \hline
Dallas       & 0.33  & 0.37   & 0.4    & 0.43   & 0.84  & 0.83    & 0.73   & 0.73   & 0.43  & 0.47    & 0.47   & 0.45   \\ \hline
SanDiego     & 0.06  & 0.16   & 0.01   & 0.23   & 0.99  & 0.82    & 0.84   & 0.90   & 0.46  & 0.48    & 0.46   & 0.37   \\ \hline
SanJose      & 0.58  & 0.69   & 0.7    & 0.66   & 0.49  & 0.48    & 0.53   & 0.52   & 0.72  & 0.76    & 0.81   & 0.66   \\ \hline
Phoenix      & 0.64  & 0.61   & 0.64   & 0.56   & 0.36  & 0.39    & 0.42   & 0.46   & 0.79  & 0.73    & 0.75   & 0.68   \\ \hline
\end{tabular}
}
\\`NA': no data available 
\label{tab:parameter_result_alpha}
\end{table}

We also compare the results row by row to detect any changes over different cities. The parameter changes in Tables \ref{tab:parameter_result} and \ref{tab:parameter_result_alpha} reflect different local patterns. From the table we can notice that even for the same brand, the models produce very different parameters across cities, which indicate that people's visit behaviors are affected by the regional differences \citep{mckenzie2015regional}, which may link to the size and shape of a city, POI co-location patterns, and urban spatial structure \citep{kang2012intra,yue2017measurements,gao2017extracting}.

For example, the $\beta$ is the exponent of distance in the Huff models and it reveals the impact of distance decay on visit activities and we are able to compare different spatial interaction patterns using the value of $\beta$ \citep{liu2014uncovering}. In general, a larger $\beta$ means the activities are more affected by the change of distances. Usually, with more spatial interactions in a city, we can expect a smaller $\beta$ as people are less spatially separated with the support of modern multi-mode transportation \citep{liu2014uncovering,mckenzie2014access,su2017geo}. By comparing the $\beta$ changes over different cities, it is clear that New York has a very small $\beta$ for both Whole Foods and Trader Joe's compared with other cities. This indicates the POI visit patterns for people in New York are less influenced by the distance. This is reasonable as the well-developed transportation makes people in such a large metropolitan city more connected to each other and long-distance will have a less negative impact in terms of preventing people from traveling to other places. We also use the averaged $\beta$ for each city to reflect the effects of distance to cities. The top cities with smallest $\beta$ from the result Table \ref{tab:parameter_result} are New York, San Diego, Philadelphia, and Chicago. Except for San Diego which has a very small $\beta$ in its Whole Foods result (there is only one Whole Foods Market in San Diego in our dataset), the other three cities are all cities with well-developed public transit systems. The mixed mode of private-driving and public transportation may make distance less sensitive for traveling and leads to small $\beta$ for those cities. 

Next, we compare the parameter differences over different types of stores and find some distinct patterns between the supermarket \& grocery stores and the department stores. Here the supermarket \& grocery stores are represented by two brands ``Whole Foods" and ``Trader Joe's" and the department stores are represented by``Ross Stores". In the 9 cities that have ``Ross Stores", 6 of them have smaller averaged $\beta$ for ``Ross Stores" compared with that for ``Whole Foods" and ``Trader Joe's". As we showed in section \ref{sec:dist_distribution}, the department stores have a smoother distance decay slope compared with that of supermarkets and grocery stores, which means that the distance affects more for visits to supermarket and grocery stores. From our result, a majority of the cities showed the same trend that distance plays a more important role when people visit supermarkets and grocery stores. This corresponds to the daily experiences as customers tend to go to the closest supermarkets or grocery stores as the goods in those types of stores are generally similar. Therefore, the distance becomes the major factor to consider when deciding which store to visit and this is also validated by our data-driven analytical results.

\subsection{Location business insights}\label{sec:regression}
In addition to the store attraction and distance, we further conduct the multiple linear regression (MLR) analysis to discover potential factors explaining why people from certain neighborhood often go to a particular POI with regards to the characteristics of that neighborhood and the POI attraction. Specifically, we take factors from the demographic and socioeconomic aspects into consideration to detect whether people from a certain socioeconomic neighborhood will have common mobility patterns in terms of the places they often visit. The dependent variable is the pairwise visit count from a CBG community to a store and the independent variables with low-multicollinearity are \textit{store total visit counts}, \textit{distance} between a store and a customer's most frequently visited home (work) CBG, \textit{total population} of a CBG, the \textit{median age} and the \textit{median household income} of people living in that CBG, and the Shannon \textit{entropy} based on natural logarithm (Ln) to measure the race $\&$ ethnicity diversity of each CBG community \citep{shannon1948mathematical}. A higher entropy value means a higher race $\&$ ethnicity diversity while lower entropy indicates a larger portion of dominated race $\&$ ethnicity group in a CBG \citep{prestby2019understanding}. Table \ref{tab:regressioncoefficients} shows the MLR coefficients of those variables estimated from the ordinary-least-squares approach and their statistical significance for explaining the overall variability of the visit probability to three brands' stores (i.e., Whole Foods, Trader Joe's and Ross) across the ten cities. The experiments demonstrate that the store attractiveness measured by the \textit{total visit counts} (attractiveness) and the \textit{median household income} are significant positive factors that drive the visits from CBGs to the stores of all three brands. The distance plays a significant negative role for both Whole Foods and Ross Stores but not for Trader Joe's. The race and ethnicity diversity (\textit{entropy measure}) has a significant positive influence for the Ross and Trader Joe' s store visits. The \textit{median age} of people in CBGs seems not to play a significant role except for Ross Stores where all factors are significant.

Furthermore, we investigate whether the customer visit patterns to the three brands and the performance of influential factors are different across these U.S. cities. Table \ref{tab:regression_rsquared} shows the R-squared values for the three brands' store visits in the MLR models. Overall the regression models perform better in Supermarkets and Grocery (Trader Joe's mean R-squared 0.279 and Whole Foods' mean R-squared 0.235) than in Department Stores (Ross stores' mean R-squared 0.161). However, there exists large regional variability of the MLR model performance in explaining the store visit patterns. The standard deviation of R-squared for Trader Joe's (0.164) is the largest among three brands. The regression model has a higher goodness of fit for the Trader Joe's stores in Phoenix, San Diego, and New York (with all larger than 0.4 R-squared respectively) but it has a very low R-squared value in Dallas (0.07). Given the large size and socioeconomic complexity of these most populated cities, there might exist other indicative features that we need to further investigate in the future.

\begin{table}[H]
\caption{The regression coefficients of influential variables for explaining the total visit variability to three brands' stores.}
\small
\centering
\resizebox{1.3\textwidth}{!}{\begin{tabular}{|c|c|c|c|c|c|c|}
\hline
 & \multicolumn{2}{c|}{Whole Foods} & \multicolumn{2}{c|}{Trader Joe's} & \multicolumn{2}{c|}{Ross Stores} \\ \hline
& Coefficients    & Sig.           & Coefficients    & Sig.            & Coefficients    & Sig.           \\ \hline
Intercept           & 3.399e+01       & 0.0016 **      & 1.135e+00       & 0.9341          & 1.867e+01       & 1.73e-06 ***   \\ \hline
Total visit counts  & 1.323e-03       & 0.0451 *       & 4.112e-03       & 0.0007 ***      & 3.980e-03       & $<$2e-16 ***      \\ \hline
Distance            & -9.218e-01      & 1.76e-07 ***   & -1.436e-02      & 0.2943          & -5.551e-01      & $<$2e-16 ***      \\ \hline
Total population    & 8.147e-04       & 0.0592 .        & 3.739e-03       & 7.31e-06 ***    & 4.132e-03       & $<$2e-16 ***      \\ \hline
Median household income & 1.431e-04       & 7.37e-07 ***   & 1.411e-04       & 0.0001 ***      & 4.369e-05       & 0.0423 *       \\ \hline
Median age          & -2.488e-01      & 0.1609         & -2.579e-01      & 0.27144         & -3.155e-01      & 0.0014 **      \\ \hline
Entropy             & -6.418e-01      & 0.8994         & 1.567e+01       & 0.0168 *        & 7.337e+00       & 0.0002 ***     \\ \hline
\end{tabular}}
\\Significance level (p-value): 0 `***' 0.001 `**' 0.01 `*' 0.05 `.' 0.1 
\label{tab:regressioncoefficients}
\end{table}

\begin{table}[H]
\caption{The R-squared for the regression models of three brands across the most populated US cities.}
\small
\centering
\begin{tabular}{|c|c|c|c|}
\hline
City  & Whole Foods & Trader Joe's & Ross Stores \\ \hline
Los Angeles  & 0.265       & 0.168        & 0.242       \\ \hline
Houston      & 0.096       & 0.119        & 0.101       \\ \hline
Chicago      & 0.131       & 0.391        & 0.089       \\ \hline
Philadelphia & 0.289       & 0.187        & 0.101       \\ \hline
New York     & 0.293       & 0.431        & NA          \\ \hline
San Antonio  & 0.381       & 0.202        & 0.126       \\ \hline
Dallas       & 0.272       & 0.070        & 0.203       \\ \hline
San Diego    & 0.240       & 0.436        & 0.224       \\ \hline
San Jose     & 0.165       & 0.222        & 0.205       \\ \hline
Phoenix      & 0.222       & 0.567        & 0.160       \\ \hline
\textbf{R-squared Mean}      & 0.235       & 0.279        & 0.161       \\ \hline
\textbf{R-squared Std.}      & 0.085       & 0.164        & 0.059       \\ \hline
\end{tabular}
\\`NA': no data available 
\label{tab:regression_rsquared}
\end{table}

\section{Conclusion and future work} \label{sec:conclusion}
In this research, we present a time-aware dynamic Huff model (T-Huff) that incorporates the hourly temporal variability of store visits to delineate the dynamic trade areas for different types of business POIs. To calibrate the model parameters, we apply the PSO technique with hourly POI visit probability derived from a  large-scale mobile phone location data set across ten most populated U.S. cities. To answer the two research questions that we posed at the beginning of this research: 
\par RQ 1: The calibrated dynamic Huff model (T-Huff) is more accurate than the original static Huff model without temporal variation consideration in predicting the market share of different types of business (e.g., supermarkets vs. department stores) over time. 

\par RQ 2: The spatial proximity, demographic and socioeconomic factors (e.g., median household income) have significant impacts on the customer choice of particular store visits. There exists regional variability for store visit patterns across different cities with varying calibrated Huff model parameters and different goodness of fit values in MLR models. The performance variability of models may link to different spatial socioeconomic structure and transportation infrastructure in those large cities. 

In sum, the presented time-aware dynamic Hull models and the analytical workflow using location big data can be applied for other categories of business stores for location-based marketing and dynamic trade area analyses.  

One limitation of our current analysis is the lack of the street-network distance and centrality measures that may influence the spatial distribution of business stores \citep{porta2009street}. In addition, the travel time and traffic congestion contexts for certain routes to store visits in human's minds may also impact their accessibility and decision makings \citep{stanley1976image,mckenzie2014access,su2017geo}. We will consider street-network measures and traffic information into the modeling framework in the future work. 

\section*{Acknowledgement}
We would like to thank the SafeGraph Inc. for providing the anonymous location visit data support. Support for this research was provided by the University of Wisconsin - Madison Office of the Vice Chancellor for Research and Graduate Education with funding from the Wisconsin Alumni Research Foundation. This research was also supported by The We Company (WeWork) through academic collaboration with the Geospatial Data Science (GeoDS) Lab at the University of Wisconsin - Madison. 

\bibliographystyle{harvard}
\bibliography{references}

\end{document}